\documentstyle [fleqn,12pt] {article}
\begin{document}
\pagestyle{empty}
\begin{flushleft}
\end{flushleft}
\begin{flushright}
RUP-96-7 \\
YITP Workshop
\end{flushright}

\vspace{5mm}
\begin{center}
{\bf Thermofield Dynamics of the Closed Bosonic String \\
--- Physical Aspects of the Thermal Duality ---}

\vspace{1cm} 
H. Fujisaki$^*$ 
and K.Nakagawa$^{**}$

\vspace{5mm}

$^*$ {\it Department of Physics, Rikkyo University, Tokyo 171}\\
$^{**}$ {\it Faculty of Pharmaceutical Sciences, Hoshi University, Tokyo 142}\\
\vspace*{2cm}
{\bf ABSTRACT}
\end{center}

\indent The thermofield dynamics of the $D  = 26$  closed bosonic thermal string 
theory is described in proper reference to the thermal duality symmetry as well 
as the thermal stability of modular invariance in association with the global 
phase structure of the bosonic thermal string ensemble. 

\vspace*{2cm}
\noindent To appear in {\it ^^ ^^ Proceedings of the YITP Workshop on Thermal 
Field Theories and their applications'', Kyoto, August 
26-28, 1996}.  

\newpage
%%%%%%%%%%%%%%%%%
%%%%%%%%%%%%%%%%%
\pagestyle{plain}
\setcounter{page}{1}
\indent Building up thermal string theories based upon the thermofield 
dynamics (TFD) has gradually been endeavoured in leaps and bounds. In the present 
communication, the TFD algorithm of the  $D = 26$  closed 
bosonic thermal string theory is commentarially exemplified  
$\grave{a}\: la$  recent papers of ourselves \cite{fujisaki1}, \cite{fujisaki2} 
through the one-loop mass shift of the dilaton, graviton and  antisymmetric 
tensor particle with proper reference to the infrared behaviour of the 
one-loop thermal cosmological constant.  Physical aspects of the TFD thermal 
string  amplitude are then described in connection with the global phase 
structure of the thermal string ensemble. 

Let us start with the one-loop self-energy amplitude $A(k_1; \zeta_1, 
\zeta_2; \beta)$ of the massless thermal tensor boson as follows:  

\begin{equation}
A(k_1; \zeta_1, \zeta_2; \beta) = - i\kappa^2 \int^{\infty}_{-\infty} 
d^Dp \; {\rm Tr}\left[ {\mit\Delta}^\beta (p) V(k_1; \zeta_1, 
\overline{\zeta}_1) 
{\mit\Delta}^\beta (p) V(k_2; \zeta_2, \overline{\zeta}_2) \right]
\end{equation}

\vspace{5mm}
\noindent at any finite temperature in the $D = 26$ closed bosonic 
thermal string theory based upon the TFD algorithm, where $\kappa, 
p^\mu, k_{r}^{\mu}; r = 1, 2$ and $\zeta_{r}^{\mu \nu} = \zeta_{r}^{\mu}
\bar{\zeta}_{r}^{\nu}; r = 1,2$ read the coupling constant, loop 
momentum, external momenta and polarization tensors, respectively, and 
$V(k; \zeta, \bar{\zeta})$ is referred to as the vertex for the 
emission or absorption of the massless tensor boson.  Here the thermal 
propagator ${\mit\Delta}^\beta (p)$ of the free closed bosonic string is 
written $\grave{a}\: la$ Leblanc in the form 

\begin{eqnarray}
{\mit\Delta}^\beta (p) & = & \int^{\pi}_{-\pi} \frac{d\phi}{4\pi}\; e^{i\phi (L_0 - 
\bar{L}_0)} \Biggl[ \int^{1}_{0} dx \: x^{L_0+\bar{L}_0-2\alpha -1} \nonumber \\
& & \mbox{ } + \frac{1}{2} \sum^{\infty}_{n=0} \frac{\delta [\alpha^\prime/2 \cdot p^2 
+ 2(n - \alpha)]}{e^{\beta |p_0|} - 1} \; \oint_c d\hat{x}\: \hat{x}^{L_0+
\bar{L}_0-2\alpha -1} \Biggr]\rule{0mm}{1cm} \quad, 
\end{eqnarray}

\vspace{5mm}
\noindent where $\stackrel{[\normalsize{-}]}{L}_0 = \alpha^\prime /4 
\cdot p^2 + \stackrel{[\normalsize{-}]}{N}$ in which 
$\stackrel{[\normalsize{-}]}{N}$ reads the number operator of the 
right-[left-]moving mode, the slope and intercept of the closed string 
reggeon are $\alpha^\prime /2$ and $2\alpha = (D - 2)/12$, 
respectively and the contour $c$ is taken as the unit circle around the 
origin.  We are then led to $A(k_1; \zeta_1, \zeta_2; \beta) = A(k_1; 
\zeta_1, \zeta_2) + \bar{A}(k_1; \zeta_1, \zeta_2; \beta)$ at any value 
of $\beta$, where use has been made of $k_{1}^{\mu} + k_{2}^{\mu} = 0$ and 
$k_{1}^{2} = k_{2}^{2} = k_1 \cdot \zeta_1 = k_1 \cdot \bar{\zeta}_1 = 
k_2 \cdot \zeta_2 = k_2 \cdot \bar{\zeta}_2 = 0$.  The  $D = 26$  
zero-temperature amplitude $A(k_1; \zeta_1, \zeta_2)$ is written $\grave{a}\: 
la$ Panda in the modular invariant fashion as follows:

\begin{eqnarray}    
\lefteqn{A(k_1; \zeta_1, \zeta_2) = (\pi \kappa)^2 (\alpha^\prime)^{-D/2} 
\; \zeta_{1}^{\mu \nu} \zeta_{2}^{\sigma \rho} \int_{F} d^2\tau \int_{P} 
d^2\nu \, \tau_{2}^{-D/2}} \nonumber \\
& & \times \exp \left[ \pi \tau_2 \cdot \frac{D - 2}{6}\right] \left| 
f(e^{2\pi i \tau}) \right| ^{-2(D-2)} \rule{0mm}{1cm} \nonumber \\
& & \times \left[ \left( \frac{\alpha^\prime}{8\pi \tau_2} \right) ^2 
(\eta_{\mu \nu} \eta_{\sigma \rho} + \eta_{\mu \rho} \eta_{\nu \sigma}) + 
\left( \frac{\alpha^\prime}{8\pi^2} \right) ^2 \eta_{\mu \sigma} \eta_{\nu \rho} 
 \left| \frac{\pi}{\tau_2} + \frac{\partial}{\partial \nu} \left\{ 
 \frac{\vartheta_{1}^{\prime} (\nu - \tau | \tau)}{\vartheta_1 (\nu - \tau | \tau)} 
 \right\} \right| ^2 \right] ,\rule{0mm}{1cm}
\end{eqnarray} 

\vspace{5mm}
\noindent where $f(w) = \prod_{n=1}^{\infty} (1 - w^n)\; ; w = q^2 = 
e^{2\pi i \tau}$, $\eta_{\kappa \tau}$ is the space-time metric, $\vartheta_1$ reads 
the Jacobi theta function, $F$  denotes the fundamental domain of the modular group 
$SL (2, Z)$ in the complex $\tau$ plane and the integration over the complex $\nu$ plane 
is restricted to cover a single parallelogrammatic region $P$.  Since the soft domain 
$k_1 \simeq 0$ is necessary and 
sufficient at any finite temperature for the dynamical mass shift of the 
massless thermal tensor boson, the present discussion is confined to the 
asymptotic behaviour of the $D  = 26$  temperature-dependent amplitude 
$\bar{A}(k_1 ; \zeta_1, \zeta_2; \beta)$ at the low energy limit $k_{10} 
\simeq 0$ .  We are then eventually led to the ^^ ^^ proper-time'' integral representation 
of $\bar{A}(k_1 ; \zeta_1, \zeta_2; \beta)$ as follows: 

\begin{eqnarray}
\bar{A}(k_1 ; \zeta_1, \zeta_2; \beta) & = & 4\pi 
\left( \frac{\kappa}{4\pi} \right) ^2 (\alpha^\prime)^{-D/2} 
\int_{0}^{\infty} d\tau_2 \, \tau_2^{-D/2} \exp \left[ \pi \tau_2 \cdot 
\frac{D - 2}{6} \right] \nonumber \\ 
& & \times \int_{-\pi}^{\pi} d\phi_1 \int_{-\pi}^{\pi} d\phi_2 \int_{0}^{1} 
\frac{dx_1}{x_1} \, \theta (x_1 - e^{-2\pi \tau_2}) \rule{0mm}{1cm} \nonumber \\
& & \times \sum_{\ell=1}^{\infty} \exp \left[ - \frac{\sigma \ell^2 
\beta^2}{2\tau_2} \right] F (\tau_2 ; z_2, \bar{z}_2, z_1z_2, \bar{z}_1\bar{z}_2) 
\rule{0mm}{1cm} \nonumber \\
& & \mbox{} + i\pi^2 \left( \frac{\kappa}{4\pi} \right) ^2 
(\alpha^\prime)^{-D/2} \int_{0}^{\infty} d\tau_2 \, \tau_{2}^{-D/2} \exp 
\left[ \pi \tau_2 \cdot \frac{D - 2}{6} \right] \rule{0mm}{1cm} \nonumber \\
& & \times \int_{-\pi}^{\pi} d\phi_1 \int_{-\pi}^{\pi} d\phi_2 \sum_{\ell=1}^{\infty} 
\ell \exp \left[ - \frac{\sigma (\ell + 1)^2 \beta^2}{2\tau_2} \right] 
\theta(-|w|) \rule{0mm}{1cm} \nonumber \\
& & \times \biggl\{ F \left( \tau_2 ; 0, 0, |w| \, e^{i(\phi_1+\phi_2)}, |w| 
\, e^{-i(\phi_1+\phi_2)}\right) \rule{0mm}{1cm} \nonumber \\
& & \mbox{} + F \left( \tau_2 ; 0, 0, -|w| \, e^{i(\phi_1+\phi_2)}, -|w| 
\, e^{-i(\phi_1+\phi_2)}\right) \biggr\} \rule{0mm}{1cm} 
\end{eqnarray}

\vspace{5mm}
\noindent at $k_{10} \simeq 0$, where $z_r = x_r \exp (i\phi_r)\; ; r = 1, 
2, |w| = e^{-2\pi \tau_2}$ and $\theta$  is the step function.  We do not go into 
details of the real function $F$ but merely refer to ref. 
\cite{fujisaki1} and ref. \cite{fujisaki2}.  It is of interest to note 
that the zero-energy 
thermal amplitude ${\rm Im} \bar{A} (k_1 \simeq 0 ; \zeta_1, \zeta_2 ; \beta)$ vanishes 
identically at  $D = 26$  for any finite temperature because 
of $|w| > 0$.  

All we have to do is now reduced to carry out the regularization of the  $D = 26$  
thermal amplitude $A(k_1 ; \zeta_1, \zeta_2 ; \beta)$ at $k_{10} \simeq 0$.  The  
$D = 26$  one-loop mass shift $A(k_1 \simeq 0 ; \zeta_1, \zeta_2 ; 
\beta)$ of the dilaton, graviton and antisymmetric tensor boson is then described at 
any finite temperature in the standard fashion which is manifestly free 
of ultraviolet divergences at $\tau_2 \sim 0$ and $|\nu | \sim \infty$ for any value 
of $\beta$ and  $D$  due to modular invariance and double periodicity.  The standard 
integral representation thus obtained is still annoyed with infrared divergences near 
the endpoints $\tau_2 \sim \infty$ and $|\nu| \sim 0$, however, unless  $D < 2$ .  The 
regularization of the $\nu$ integration has already been brought to realization in the 
modular invariant fashion.  Moreover, the infrared 
divergence of the one-loop TFD self-energy amplitude at $\tau_2 \sim \infty$ can be 
remedied at any finite temperature through the dimensional regularization in the sense 
of analytic continuation which is of course modular invariant as well as double periodic.  
The dimensionally regularized, $D = 26$  one-loop dual symmetric mass 
shift $\hat{A} (k_1 \simeq 0 ; \zeta_1, \zeta_2 ; \beta)$ of the dilaton, graviton and 
antisymmetric tensor boson is then reduced to 

\begin{eqnarray}
\hat{A} (k_1 \simeq 0 ; \zeta_1, \zeta_2 ; \beta) & = & -4\pi (\pi 
\kappa)^2 (4\pi^2)^{(D-1)/2}\;  \zeta_{1}^{\mu \nu} \zeta_{2}^{\sigma \rho} 
\hat{\mit\Lambda} (\beta) \nonumber \\
& & \times \left\{ \tilde{D}_{\mu \nu \sigma \rho} + \tilde{G}_{\mu \nu \sigma \rho} + 
\tilde{T}_{\mu \nu \sigma \rho} \right\} \quad, \rule{0mm}{1cm}
\end{eqnarray}

\vspace{5mm}
\noindent where $\tilde{D}_{\mu \nu \sigma \rho} = (\alpha^\prime 
/8\pi)^2 \eta_{\mu \nu} \eta_{\sigma \rho}$ , $\tilde{G}_{\mu \nu \sigma 
\rho} = 0$, $\tilde{T}_{\mu \nu \sigma \rho} = (\alpha^\prime 
/8\pi)^2 (\eta_{\mu \rho}\eta_{\nu \sigma} - \eta_{\mu \sigma}\eta_{\nu 
\rho})$ , and $\hat{\mit\Lambda}(\beta)$ reads the dimensionally 
regularized, $D = 26$ one-loop dual symmetric thermal cosmological constant as follows;  

\begin{eqnarray}
\lefteqn{\hat{\mit\Lambda} (\beta) = - \frac{\sqrt{\pi \alpha^\prime}}{\beta} 
\: (4\pi \alpha^\prime)^{-D/2} \sum_{m,n \in Z} 
\int^{\frac{1}{2}}_{-\frac{1}{2}} d\tau_1 \; \exp [2\pi imn \tau_1]}
\nonumber \\
& & \times \left( \frac{\beta^2}{4\pi^2 \alpha^\prime} \: m^2 + \frac{4\pi^2 
\alpha^\prime}{\beta^2} \: n^2 - \frac{D - 2}{6} \right)^{(D-1)/2}
\rule{0mm}{1cm} \nonumber \\
& & \times {\mit\Gamma} \left[ - \frac{D - 1}{2}\: , \pi \sqrt{1 - 
\tau_1^{2}} \left( 
\frac{\beta^2}{4\pi^2 \alpha^\prime} \: m^2 + \frac{4\pi^2 
\alpha^\prime}{\beta^2} \: n^2 - \frac{D - 2}{6} \right) \right] . 
\rule{0mm}{1cm}
\label{eq:hatLambda}
\end{eqnarray} 

\vspace{5mm}
\noindent It is a matter of course that $\tilde{D}_{\mu \nu \sigma \rho}, 
\tilde{G}_{\mu \nu \sigma \rho}$ and  
$\tilde{T}_{\mu \nu \sigma \rho}$ describe the factors of the dilaton, graviton and 
antisymmetric tensor boson contribution, respectively, to the one-loop thermal amplitude 
$\hat{A} (k_1 \simeq 0 ; \zeta_1, \zeta_2 ; \beta)$.  The thermal duality symmetry 
$\beta \hat{A} (k_1 \simeq 0 ; \zeta_1, \zeta_2 ; \beta) = \tilde{\beta} \hat{A} (k_1 
\simeq 0 ; \zeta_1, \zeta_2 ; \tilde{\beta})$ then follows for any value 
of  $\beta$ as a simple and natural consequence of the thermal duality 
relation $\beta \hat{\mit\Lambda}(\beta) = \tilde{\beta} \hat{\mit\Lambda} 
(\tilde{\beta})$, 
where $\tilde{\beta} = 4\pi^2 \alpha^\prime /\beta$.  In accordance, the dimensionally 
regularized, thermal amplitude $\hat{A} (k_1 \simeq 0 ; \zeta_1, \zeta_2 ; \beta)$ yields 
the non-vanishing one-loop dual symmetric mass shift for the dilaton and antisymmetric 
tensor boson which is literally proportional at any finite temperature to the 
dimensionally regularized, $D  = 26$  one-loop dual symmetric thermal cosmological 
constant $\hat{\mit\Lambda} (\beta)$.  The dimensionally regularized, one-loop 
dual symmetric mass shift of the graviton on the other hand is of course guaranteed to 
vanish identically at any finite temperature.  It will be possible to argue that these 
observations based upon the TFD paradigm are in full consonance with the thermal 
stability of renormalizability, factorizability, duality and gauge invariance which is 
in turn substantiated at the soft limit $k_1 \simeq 0$ as an immediate consequence of the 
thermal stability of both modular invariance and double periodicity.  
It is parenthetically mentioned that the so-called $i\varepsilon$ 
prescription $D + i\varepsilon$ is adopted under explicit evaluation 
of the dimensionally regularized, $D = 26$ thermal amplitude 
$\hat{\mit\Lambda}(\beta)$ in full agreement with the off-shell 
continuation algorithm which inevitably yields finite but complex 
values for $\hat{\mit\Lambda}(\beta)$ in direct association with the 
nonvanishing decay rate of the false tachyonic [thermal] vacuum.  For 
the detailed paradigm, we merely refer to ref. \cite{fujisaki1}.  

Let us examine the singularity structure of the dimensionally 
regularized, $D = 26$  one-loop dual 
symmetric free energy amplitude $\hat{\mit\Lambda}(\beta)$.  The 
position of the singularity  $\beta_{|m|, |n|}$ is determined by solving 
$\beta /\tilde{\beta} \cdot m^2 + \tilde{\beta} /\beta \cdot n^2 - 4 = 0$ for every 
allowed $(m, n)$ in eq.($\!$$\!$ ~\ref{eq:hatLambda}).   
In particular, $\beta_{1, 0}$ and $\tilde{\beta}_{1, 0}$ form the leading branch points at 
$\beta_H = 4\pi\sqrt{\alpha^\prime}$ and $\tilde{\beta}_H = \pi \sqrt{\alpha^\prime}$, 
respectively, where $\beta_H \; [\tilde{\beta}_H]$ reads the inverse [dual] Hagedorn 
temperature.  Moreover, there appears the 
self-dual leading branch point at $\beta_{2, 0} = \tilde{\beta}_{2, 0} = \beta_0 
= 2\pi \sqrt{\alpha^\prime}$ as an inevitable consequence of the thermal 
duality symmetry.  In addition, $\beta_{1, 1}$ and $\tilde{\beta}_{1, 1}$  yield the 
non-leading branch points at $(\sqrt{3} + 1) \, \pi \sqrt{2\alpha^\prime}$ and 
$(\sqrt{3} - 1) \, \pi \sqrt{2\alpha^\prime}$, respectively.  Finally, all the residual 
secondary branch points at $\beta_{|m|, 0} \; [\tilde{\beta}_{|m|, 0}]$ with 
$m = \pm 3; \pm 4; \cdots$  are, of course, removed onto the unphysical sheet of the 
physical $\tilde{\beta} \; [\beta]$ channel across the leading branch cut mentioned above.  
Let us now turn our attention to the statistical ensemble of the $D  = 26$  closed 
bosonic thermal string.  The thermodynamical properties of the bosonic thermal string 
excitation can be analyzed through the microcanonical ensemble paradigm outside the 
analyticity domain of the canonical ensemble.  Substantial use is made of the thermal 
duality relation not only for the canonical region but also for the microcanonical region.  
There will then exist four phases as follows:  I) the 
$\beta$ channel canonical phase in the range $4\pi \sqrt{\alpha^\prime} = 
\beta_H \leq \beta < \infty$, II) the dual $\tilde{\beta}$ channel canonical phase in the 
range $0 < \beta \leq \tilde{\beta}_H = \beta_H/4$,   
III) the $\beta$ channel microcanonical domain $\beta_H /2 = \beta_0 
\leq \beta < \beta_H$ and IV) the dual $\tilde{\beta}$ channel microcanonical domain 
$\tilde{\beta}_H < \beta \leq \tilde{\beta}_0 = \beta_0$.  The present observation might lead up to a novel 
hypothesis of the ^^ ^^ extended" $\beta$ [dual $\tilde{\beta}$] channel microcanonical phase 
for the range $\beta_0 \leq \beta < \beta_{1,1} \; [\tilde{\beta}_{1,1} < 
\beta \leq \tilde{\beta}_0 = \beta_0]$ .  Another newfangled hypothesis of the $\beta$ 
[dual $\tilde{\beta}$] channel microcanonical phase confined to the 
region $\beta_{1,1} \leq \beta < \beta_H \; [\tilde{\beta}_H < \beta \leq 
\tilde{\beta}_{1,1}]$  might not be abandoned yet, however.  The 
future exploration of the thermodynamical properties of string excitations is 
inevitable for the manifest materialization of the physical significance of $\beta_{1,1}$ 
as well as $\tilde{\beta}_{1,1}$, anyhow.  It will be possible to 
claim that we have succeeded in shedding some light upon the global 
phase structure of the thermal string ensemble in proper reference to 
the thermal duality symmetry for the $D = 26$ closed bosonic thermal 
string theory based upon the TFD algorithm.    

Let us conclude by emphasizing that the present TFD paradigm will deserve more than 
passing consideration in the thermodynamical investigation of the thermal string 
excitation in general.

Last but not least, one of the authors (H. F.) cherishes warm 
encouragement of the late lamented Professor H. Umezawa with deep 
gratitude.

%%%%%%%%
%%%%%%%%

\end{document}